\documentclass[english,superscriptaddress,letterpaper,aps,prb,twocolumn]{revtex4-2}
\usepackage{lmodern}
\usepackage[LGR,T1]{fontenc}
\usepackage[latin1]{inputenc}
\usepackage{color}
\usepackage{dsfont}
\usepackage{amsmath}
\usepackage{amssymb}
\usepackage{graphicx}

\makeatletter

\DeclareRobustCommand{\greektext}{%
  \fontencoding{LGR}\selectfont\def\encodingdefault{LGR}}
\DeclareRobustCommand{\textgreek}[1]{\leavevmode{\greektext #1}}

\@ifundefined{textcolor}{}
{%
 \definecolor{BLACK}{gray}{0}
 \definecolor{WHITE}{gray}{1}
 \definecolor{RED}{rgb}{1,0,0}
 \definecolor{GREEN}{rgb}{0,1,0}
 \definecolor{BLUE}{rgb}{0,0,1}
 \definecolor{CYAN}{cmyk}{1,0,0,0}
 \definecolor{MAGENTA}{cmyk}{0,1,0,0}
 \definecolor{YELLOW}{cmyk}{0,0,1,0}
}

\usepackage{lmodern}

\makeatletter

\usepackage{lmodern}

\makeatletter
\usepackage{lmodern}

\makeatletter

\makeatletter

\usepackage{epsfig}

\usepackage{dcolumn}

\usepackage{bm}

\usepackage{bbm}

\usepackage{wasysym}

\usepackage{marvosym}

\usepackage{subfloat}% need for subequations
\usepackage{color}% use if color is used in text
\usepackage{subfigure}% use for side-by-side figures

\usepackage{blindtext}

\newlength{\textwidthm}

\makeatother

\makeatother

\usepackage{tikz}

\makeatother

\makeatother

\makeatother

\usepackage{babel}
\begin{document}
\title{Thermal and thermoelectric transport in flat bands with non-trivial quantum geometry}

\author{Kevin Wen}
\affiliation{Department of Physics, University of Texas at Austin, Austin, Texas 78712, USA}
\author{Hong-Yi Xie}
\affiliation{Department of Physics and Astronomy, University of Oklahoma, Norman, OK 73069, USA}
\author{Assa Auerbach}
\affiliation{Physics Department, Technion, 32000 Haifa, Israel}
\author{Bruno Uchoa}  \email{uchoa@ou.edu}
\affiliation{Department of Physics and Astronomy, University of Oklahoma, Norman, OK 73069, USA}

\date{\today}
\begin{abstract}
Although quasiparticles in flat bands have zero group velocity, they can display an anomalous velocity  due to the quantum geometry. 
We address the thermal and thermoelectric transport in flat bands in the clean limit with a small amount of broadening due to inelastic scattering. 
We derive general Kubo formulas for flat bands in the DC limit up to linear order in the broadening and extract expressions for the thermal conductivity, the Seebeck and Nernst coefficients.
We show that the Seebeck coefficient for flat Chern bands is topological up to second order corrections in the broadening. 
We identify thermal and thermoelectric transport signatures for two generic flat Chern bands and also for the generalized flattened Lieb model, which describes a family of three equally spaced flat Chern bands where the middle one is topologically trivial. 
Finally, we address the saturation of the quantum metric lower bound for a general family of Hamiltonians with an arbitrary number of flat Chern bands corresponding to SU(2)
coherent states. 
We find that only the extremal bands in this class of Hamiltonians saturate the bound, provided that the momentum dependence of their Hamiltonians is described by a meromorphic function. 
\end{abstract}

\maketitle

\section{Introduction}

By endowing the Hilbert space with a metric and a curvature, the modern theory of solids resorts to tools from differential geometry and topology
to analyze the physical properties of electrons in a crystal~\cite{Thouless1982,Haldane1988,Berry1984,Mele2005,Provost1980,Xiao}.
If $\langle\mathbf{r}|u_{n,\mathbf{k}}\rangle\equiv\langle\mathbf{r}|n\rangle$ is the periodic part of the Bloch wavefunction for a band labeled by index $n$, the quantum geometric tensor is generally defined as
\cite{Vanderbilt}
\begin{align}
\mathcal{Q}_{n}^{\gamma\delta} & =\langle\partial_{\gamma}n|\left[1-|n\rangle\langle n|\right]|\partial_{\delta}n\rangle\equiv g_{n}^{\gamma\delta}+\frac{i}{2}\varepsilon^{\gamma\delta}\Omega_{n},\label{eq:gij}
\end{align}
where $\gamma,\delta=x,y$ are the directional indices, $\varepsilon^{\gamma\delta}$ is the antisymmetric tensor and $\partial_{\gamma}$ is the crystal momentum derivative $\partial/\partial k_{\gamma}$, with $|\partial_{\gamma}n\rangle\equiv\partial_{\gamma}|u_{n,\mathbf{k}}\rangle$.
The symmetric part of the quantum geometric tensor $g_{n}^{ab}$ is the quantum metric, also known as the Study-Fubini metric, whereas the antisymmetric part $\Omega_{n}^{\gamma\delta}\equiv\varepsilon^{\gamma\delta}\Omega_{n}$ is the Berry curvature, whose integral over the Brillouin zone (BZ) gives the Chern number of the $n$-th band~\cite{Thouless1982}. 

Manifestations of the quantum geometry are generally believed to be more prominent in flat bands, where the mass of the quasiparticles becomes infinite. 
In the flat band limit, the superfluid weight was predicted to have a lower bound set by the Chern number~\cite{Peotta,Julku,Torma2}.
Quantum geometric effects are credited to the presence of superfluid weight anomalies in the flat bands of twisted graphene bilayer~\cite{Tian} and to the presence of Lamb shifts in the excitonic spectrum of dichalchogenide materials~\cite{Srivastava2015,Zhou}. Very recently, it has been predicted the existence of topological excitons~\cite{Xie}, whose
profile wave function has a finite vorticity due to a combination of topological and quantum geometric effects in the electronic bands. 

Transport in perfectly flat bands is ruled by the properties of the quantum geometry. Whereas transverse conductivities follow from the antisymmetric part of the quantum geometric tensor, as in the conventional integer quantum Hall effect, the longitudinal response is primarily determined by the quantum metric. 
Writing a generic Hamiltonian $\hat{h}_{\mathbf{k}}=U_{\mathbf{k}}^{\dagger}\hat{\varepsilon}_{\mathbf{k}}U_{\mathbf{k}}$, where $U$ is some unitary transformation that diagonalizes $\hat{h}_{\mathbf{k}}$ and $\hat{\varepsilon}_{\mathbf{k}}=\text{diag}[E_{n}]$ is diagonal, the velocity operator of the quasiparticles in the diagonal basis is (we set $\hbar\to 1$)
\begin{equation}
\hat{v}_{\gamma,\mathbf{k}}^{\text{d}}=\partial_{\gamma}\hat{\varepsilon}_{\mathbf{k}}+[\hat{\mathcal{A}}_{\gamma,\mathbf{k}},\hat{\varepsilon}_{\mathbf{k}}],\label{eq:vel}
\end{equation}
where $\hat{\mathcal{A}}_{\gamma,\mathbf{k}}=U_{\mathbf{k}}\partial_{\gamma}U_{\mathbf{k}}^{\dagger}$ is the Berry dipole tensor \cite{Haldane}. In flat bands, which are
dispersionless, the group velocity $\partial_{\gamma}\hat{\varepsilon}_{\mathbf{k}}=0$, whereas the commutator in the second term gives an anomalous contribution to the quasiparticle velocity. This term follows from the off-diagonal components of the Berry dipole tensor $\hat{\mathcal{A}}_{\gamma,\mathbf{k}}$, reflecting only interband contributions, $\langle n|\hat{v}_{\gamma}^{\text{d}}|m\rangle=(E_{m}-E_{n})\langle n|\hat{\mathcal{A}}_{\gamma,\mathbf{k}}|m\rangle$.
Even though perfectly flat bands lack a Fermi surface, in the presence of inelastic scattering with the lattice and interaction effects, they can display finite DC longitudinal quantum transport and possibly other physical properties commonly observed in metals~\cite{Torma}. 

In this work, we address the thermal and thermoelectric transport of electronic flat bands with non-trivial quantum geometry. 
The thermoelectric response is characterized in experiments by the Seebeck and Nernst coefficients, which respectively measure longitudinal and transverse
electric fields resulting from the application of a temperature gradient.
We consider the clean limit and assume that the bands have a small amount of broadening $\eta$ due to inelastic processes. 
Those processes are required to cool the system and avoid Joule heating effects~\cite{Tremblay}.
In Chern bands, the broadening permits the simultaneous presence of
finite longitudinal and transverse transport coefficients, while promoting
transitions among the flat bands in the free particle limit. We derive
the general Kubo formulas for flat bands in the DC limit to linear
order in the broadening and the corresponding expressions for the
Nernst and Seebeck coefficients. 

We examine the lower bound saturation of the trace of the quantum
metric for a general family of Hamiltonians with SU(2) coherent eigenstates
that describe an arbitrary number of equally spaced flat bands. We
show that in this family of Hamiltonians only the extremal bands in
the energy spectrum saturate the lower bound, provided that the momentum
dependence of the Hamiltonian is described by a meromorphic function.
We then explore the heat transport of two sub-classes of Hamiltonians
with flat bands: the case of two arbitrary flat Chern bands with opposite
Chern numbers and a generalized flattened Lieb model, which describes
any three equally spaced flat bands, where the two outer ones are
topological, with opposite Chern numbers, and the middle one is trivial.
We find that the leading contribution of the Seebeck coefficient in
flat Chern bands is of topological origin and is independent of the
broadening, whereas the Nernst coefficient is proportional to the
broadening in leading order. 

The structure of the paper is as follows. In section II we outline
the Kubo formulas for longitudinal and transverse DC transport in
flat bands using the Lehmann representation. We then address thermal
and thermoelectric transport in Chern bands under the constraint of
zero particle flow, which is of relevance to experiments. In section
III we analytically calculate the quantum metric and analyze the thermal
and thermoelectric transport for two generic flat Chern bands. We
calculate the Seebeck and Nernst coefficients and the thermal conductivity
as a function of the temperature and filling of the bands. In section
IV we generalize our analysis of the saturation of the quantum metric
bound for a family of flat band Hamiltonians constructed with SU(2)
spin coherent eigenstates. We then analytically calculate the quantum
metric of the generalized flattened Lieb model and derive the corresponding
conductivities. Finally, in section V we present a discussion of our
results. 

\section{Kubo Formulas }

The zero-momentum particle and energy current density operators are
derived from the continuity equation for charge and energy densities
respectively \cite{Paul},
\begin{align}
j_{\gamma}^{P} & =\int_{\text{BZ}}\hat{\psi}_{\mathbf{k}}^{\dagger}\hat{v}_{\gamma,\mathbf{k}}\hat{\psi}_{\mathbf{k}}\\
j_{\gamma}^{E} & =\frac{1}{2}\int_{\text{BZ}}\hat{\psi}_{\mathbf{k}}^{\dagger}\left[\hat{h}_{\mathbf{k}}\hat{v}_{\gamma,\mathbf{k}}+\hat{v}_{\gamma,\mathbf{k}}\hat{h}_{\mathbf{k}}\right]\hat{\psi}_{\mathbf{k}},\label{JQ}
\end{align}
where $\hat{h}_{\mathbf{k}}$ is a matrix that corresponds to the
momentum representation of some generic tight-binding Hamiltonian,
\begin{equation}
\mathcal{H}=\int_{\text{BZ}}\sum_{\alpha\beta}\psi_{\alpha,\mathbf{k}}^{\dagger}\left(h_{\alpha\beta,\mathbf{k}}-\mu\delta_{\alpha\beta}\right)\psi_{\beta,\mathbf{k}},\label{eq:BZ}
\end{equation}
with $\int_{\text{BZ}}\to(2\pi)^{-2}\int\text{d}^{2}k$ representing
integration over the Brillouin zone (BZ), and $\gamma=x,y$ labeling
the directions in momentum space. $\psi_{\alpha,\mathbf{k}}$ is the
annihilation operator of an electron with orbital index $\alpha=1,\dots,N$
and momentum $\mathbf{k}$, and $\hat{\psi}_{\mathbf{k}}$ the corresponding
$N$-component spinor. The matrix $\hat{v}_{\gamma,\mathbf{k}}\equiv\partial_{\gamma}\hat{h}_{\mathbf{k}}$
is the velocity operator in the orbital basis, which can be non-zero
even in the flat band limit. This operator relates to the velocity
in the diagonal basis in Eq. (\ref{eq:vel}) through a unitary transformation,
$\ensuremath{\hat{v}_{\gamma,\mathbf{k}}^{\mathrm{d}}=U_{\mathbf{k}}^{\dagger}\hat{v}_{\gamma,\mathbf{k}}U_{\mathbf{k}}}$.
The symmetrized form of the energy current density operator (\ref{JQ})
is required to ensure Hermiticity, since $\hat{v}_{\gamma,\mathbf{k}}$
does not always commute the with Hamiltonian $\hat{h}_{\mathbf{k}}$.
The heat current density $j_{\gamma}^{Q}$ and energy current density
$j_{\gamma}^{E}$ operators are related to each other by $j_{\gamma,\mathbf{q}}^{Q}=j_{\gamma,\mathbf{q}}^{E}-\mu j_{\gamma,\mathbf{q}}^{P}$,
with $\mu$ the chemical potential \cite{Mahan}. 

The real part of the finite temperature transport coefficient is \cite{Mahan,Bradlyn},
\begin{equation}
\text{Re}\left[L_{\gamma\delta}^{(AB)}(\omega_{+})\right]=\frac{1}{\omega}\text{Im}\,\Pi^{AB}(i\omega\to\omega_{+}),\label{eq:L}
\end{equation}
where 
\begin{equation}
\Pi^{AB}(i\omega)=\int_{0}^{\beta}d\tau\,e^{i\omega\tau}\big\langle T_{\tau}[j_{\gamma}^{A}(\tau)j_{\delta}^{B}(0)]\big\rangle\label{eq:Pi}
\end{equation}
is the current-current density correlation function in Matsubara frequencies,
with $T_{\tau}$ denoting the imaginary time ordered product, $\beta=1/k_{B}T$
the inverse of temperature, and $\omega_{+}=\omega+i0_{+}$. The indices
$A,B=P,Q$ label either particle or heat current operators. We note
that the imaginary part of $L_{\gamma\delta}^{(AB)}(\omega)$ follows
from application of the Kramers-Kronig relation to Eq. (\ref{eq:L}),
which is required when calculating the finite frequency response.
The electric ($\sigma$), thermoelectric ($\alpha$) and thermal ($\kappa$)
conductivity tensors are defined as (restoring $\hbar$) \cite{Mahan}
\begin{equation}
\sigma_{\gamma\delta}(\omega_{+})=\frac{e^{2}}{\hbar}\left[L_{\gamma\delta}^{(PP)}(\omega_{+})-L_{\gamma\delta}^{(PP)}(0)\right],\label{eq:sigma}
\end{equation}
\begin{equation}
\alpha_{\gamma\delta}(\omega_{+})=\frac{e}{\hbar T}\left[L_{\gamma\delta}^{(PQ)}(\omega_{+})-L_{\gamma\delta}^{(PQ)}(0)\right],\label{eq:alpha}
\end{equation}
and
\begin{equation}
\kappa_{\gamma\delta}(\omega_{+})=\frac{1}{\hbar T}\left[L_{\gamma\delta}^{(QQ)}(\omega_{+})-L_{\gamma\delta}^{(QQ)}(0)\right].\label{eq:kappa}
\end{equation}
The definitions above account for the `magnetization' subtraction
\cite{Cooper,Qin} in the particle-heat and heat-heat correlation
functions. This procedure eliminates spurious currents that are generated
by the static ``gravitational' field introduced by temperature gradients,
which are out-of-equilibrium statistical forces \cite{Luttinger,Auerbach,Chernodub}.
The subtraction eliminates spurious divergences in the thermal conductivity
that otherwise would violate the third law of thermodynamics. A detailed
account of the magnetization subtraction for flat bands is described
below in section II.B. The definition of the thermal conductivity
tensor above is valid only in the absence of experimental constraints
for particle flow, which we discuss later on.

Using the basis of eigenstates $|u_{n,\mathbf{k}}\rangle\equiv|n\rangle$
of Hamiltonian (\ref{eq:BZ}), with $n$ the band index, the clean
limit of Eq. (\ref{eq:L}) is 
\begin{equation}
\text{Re}\left[L_{\gamma\delta}^{(AB)}(\omega_{+})\right]=\frac{1}{\omega}\text{Im}\!\int_{\text{BZ}}\sum_{mn}f_{mn}\frac{j_{\gamma,mn}^{A}j_{\delta,nm}^{B}}{\omega_{mn}+\omega_{+}},\label{eq:kubo_greenwood_wick-1}
\end{equation}
where $f_{mn}\equiv f_{m}(\mathbf{k})-f_{n}(\mathbf{k})$ is the difference
between Fermi distributions in different bands, $\omega_{mn}\equiv E_{m}(\mathbf{k})-E_{n}(\mathbf{k})$
their corresponding energy difference and 
\begin{align}
j_{\gamma,mn}^{P} & \equiv-\omega_{mn}\langle m|\partial_{\gamma}n\rangle\label{eq:jP}\\
j_{\gamma,mn}^{Q} & \equiv-\frac{1}{2}(E_{m}+E_{n}-2\mu)\omega_{mn}\langle m|\partial_{\gamma}n\rangle\label{eq:jQ}
\end{align}
the matrix elements of the current operators, where $E_{m}$ is the
energy of the levels.

Thermalization requires the presence of a finite amount of broadening
$\eta>0$ due to inelastic processes involving bosonic modes, such
as phonons. The analytically continued frequency becomes $\omega_{+}=\omega+i\eta$.
We will ignore microscopic details of the bath, that are typically
encoded in the bosonic self-energy of the correlation functions, by
treating $\eta$ as a constant. This assumption is sensible in the
finite temperature regime $k_{B}T>\eta$. 

In a more explicit form, the real part of the transport coefficient
in Eq. (\ref{eq:kubo_greenwood_wick-1}) is
\begin{align}
\text{Re}\left[L_{\gamma\delta}^{(AB)}(\omega_{+})\right] & =\frac{1}{\omega}\int_{\text{BZ}}\sum_{mn}f_{mn}\times\nonumber \\
 & \negthickspace\negthickspace\negthickspace\negthickspace\negthickspace\negthickspace\negthickspace\frac{\text{Re}[j_{\gamma,mn}^{A}j_{\delta,nm}^{B}]\eta-\text{Im}[j_{\gamma,mn}^{A}j_{\delta,nm}^{B}](\omega_{mn}+\omega)}{(\omega_{mn}+\omega)^{2}+\eta^{2}}.\label{eq:L1}
\end{align}
The first term is symmetric in the $\gamma,\delta$ indices and is
related to the quantum metric, whereas the second antisymmetric term
is related to the Berry curvature. Here we consider the case of systems
that do not have a Fermi surface, such as perfectly flat bands at
any filling. Eq. (\ref{eq:L1}) is generically applicable to flat
bands. Only interband processes ($m\neq n$) contribute to the velocity
of the quasiparticles and hence to the longitudinal transport coefficients.
For the transverse part, which is non-dissipative, on-shell contributions
need to be accounted for carefully, as we show below, even in the
flat band limit. 

\subsection{Longitudinal DC conductivities}

We are interested in the DC limit of the conductivity, $\omega\to0$,
which is real. As shown in Appendix A, the DC longitudinal response
to lowest order in the broadening $\eta$ is
\begin{align}
L_{\gamma\gamma}^{(AB)}(i\eta) & =-2\eta\int_{\text{BZ}}\sum_{m\neq n}\frac{f_{mn}}{\omega_{mn}^{3}}\text{Re}[j_{\gamma,mn}^{A}j_{\gamma,nm}^{B}]+\mathcal{O}(\eta^{2}),\label{eq:ReL2-1}
\end{align}
with $\text{Im}[L_{\gamma\delta}^{(AB)}(0)]=0$. The quantum metric
and the Berry curvature can be conveniently written as a sum of their
interband matrix elements, $g_{n}^{\gamma\delta}=\sum_{m\neq n}g_{nm}^{\gamma\delta},$
and $\Omega_{n}^{\gamma\delta}=\sum_{m\neq n}\Omega_{nm}^{\gamma\delta}$,
where 
\begin{equation}
g_{nm}^{\gamma\delta}\equiv\frac{1}{2}\left(\langle\partial_{\gamma}n|m\rangle\langle m|\partial_{\delta}n\rangle+\langle\partial_{\delta}n|m\rangle\langle m|\partial_{\gamma}n\rangle\right)\label{eq:gmn}
\end{equation}
and
\begin{align}
\Omega_{nm}^{\gamma\delta} & \equiv i\big(\langle\partial_{\gamma}n|m\rangle\langle m|\partial_{\delta}n\rangle-\langle\partial_{\delta}n|m\rangle\langle m|\partial_{\gamma}n\rangle\big).\label{eq:Omegamn}
\end{align}
We arrive at general expressions for the DC longitudinal transport
coefficients in systems with flat bands to leading order in the broadening
$\eta$,
\begin{align}
L_{\gamma\gamma}^{(PP)}(i\eta) & =-2\eta\int_{\text{BZ}}\sum_{m\neq n}f_{mn}\frac{g_{mn}^{\gamma\gamma}}{\omega_{mn}}\label{eq:LPP}\\
L_{\gamma\gamma}^{(QP)}(i\eta) & =-\eta\int_{\text{BZ}}\sum_{m\neq n}f_{mn}(E_{m}+E_{n}-2\mu)\frac{g_{mn}^{\gamma\gamma}}{\omega_{mn}}\label{eq:LQP}\\
L_{\gamma\gamma}^{(QQ)}(i\eta) & =-\frac{\eta}{2}\int_{\text{BZ}}\sum_{m\neq n}f_{mn}(E_{m}+E_{n}-2\mu)^{2}\frac{g_{mn}^{\gamma\gamma}}{\omega_{mn}}.\label{eq:LQQ}
\end{align}
The magnetization subtraction plays no role in the longitudinal transport,
since clearly $L_{\gamma\gamma}^{(AB)}(0)=0$. The diagonal components
of the DC conductivity tensor $\sigma_{\gamma\gamma}(0)$ are dominated
by the \emph{interband} quantum metric integrated over the BZ and
is proportional to the broadening due to inelastic effects, which
allow for interband transitions \cite{Huhtinen,Mera,Komissarov},
\begin{equation}
\sigma_{\gamma\gamma}(0)=-\frac{2\eta e^{2}}{\hbar}\int_{\text{BZ}}\sum_{m\neq n}f_{mn}\frac{g_{mn}^{\gamma\gamma}}{\omega_{mn}}.\label{eq:sigma_long}
\end{equation}

As is well know \cite{Roy}, the trace of the quantum metric of band
$m$ for a given momentum has a local lower bound set by the Berry
curvature of the same band at the same momentum,
\begin{equation}
\text{tr}[g_{m}^{\gamma\delta}]=g_{m}^{xx}+g_{m}^{yy}\geq|\Omega_{m}^{xy}|.\label{eq:g}
\end{equation}
For two isolated bands, the same inequality also applies for the trace
of the interband quantum metric. In that case, coherent transport
in the form of a finite longitudinal conductivity in the $\omega\to0$
limit is ensured by the presence of inelastic broadening and by the
existence of a finite Berry curvature in parts of the BZ, even if
the total Chern number in each band is zero. 

The DC longitudinal thermoelectric and thermal conductivities are
(restoring $\hbar)$
\begin{equation}
\alpha_{\gamma\gamma}(i\eta)=-\frac{e\eta}{\hbar T}\int_{\text{BZ}}\sum_{m\neq n}f_{mn}(E_{m}+E_{n}-2\mu)\frac{g_{mn}^{\gamma\gamma}}{\omega_{mn}}\label{eq:alpha2-1}
\end{equation}
and 
\begin{equation}
\kappa_{\gamma\gamma}(i\eta)=-\frac{\eta}{2\hbar T}\int_{\text{BZ}}\sum_{m\neq n}f_{mn}(E_{m}+E_{n}-2\mu)^{2}\frac{g_{mn}^{\gamma\gamma}}{\omega_{mn}}.\label{eq:kappa2-1}
\end{equation}
As noted earlier, $\eta$ is related to the microscopic decay rate
of the quasiparticles due to inelastic processes and has a temperature
dependence. We note that consistency with the third law of thermodynamics
requires that the the thermal conductivity vanishes in the zero temperature
limit, and hence that $\eta(T\to0)$ goes to zero. For the purposes
of this work, we will treat $\eta$ as a parameter with an implicit
temperature dependence. 

\subsection{Transverse conductivities}

The calculation of the DC transverse transport coefficients is more
subtle than in the case of the longitudinal part because it requires
handling the magnetization subtraction \emph{before} taking the flat
band limit. From Eq. (\ref{eq:L1}), the real part of the transverse
transport coefficients in the zero frequency limit is
\begin{align}
L_{\gamma\delta}^{(AB)}(i\eta) & =\int_{\text{BZ}}\sum_{mn}f_{mn}\text{Im}[j_{\gamma,mn}^{A}j_{\delta,nm}^{B}]\nonumber \\
 & \qquad\times\left(\frac{2\omega_{mn}^{2}}{(\omega_{mn}^{2}+\eta^{2})^{2}}-\frac{1}{\omega_{mn}^{2}+\eta^{2}}\right).\label{eq:L-1}
\end{align}
Performing the magnetization subtraction and expanding to order $\eta^{4}$
in the broadening,
\begin{align}
L_{\gamma\delta}^{(AB)}(i\eta) & -L_{\gamma\delta}^{(AB)}(0)=\int_{\text{BZ}}\sum_{mn}f_{m}\nonumber \\
 & \times\text{Im}[j_{\gamma,mn}^{A},j_{\delta,nm}^{B}]\left(\frac{\eta^{2}}{\omega_{mn}^{2}+\eta^{2}}\right)+\mathcal{O}(\eta^{4}).\label{eq:deltaL}
\end{align}
where $[X,Y]$ is a commutator. The Lorentzian factor $\mathcal{L}_{mn}\equiv\eta^{2}/(\omega_{mn}+\eta^{2})$
that appears after the subtraction is responsible for the dominance
of on-shell contributions to the transverse response in the small
$\eta$ limit \cite{Auerbach}. In that limit, $\mathcal{L}_{mn}\to\delta_{mn}$
is a Kronnicker delta that keeps the summation over momenta unaffected,
but constraints the summation over the $m,n$ indices to the same
band. 

In a more explicit form, we can evaluate the commutator to express
the transverse transport coefficients after the magnetization subtraction
as
\begin{align}
L_{xy}^{(PP)}(i\eta)-L_{xy}^{(PP)}(0) & =\sum_{m}\int_{\text{BZ}}f_{m}\varepsilon_{\alpha\beta}\partial_{\alpha}A_{m,\beta}(\mathbf{k})\label{eq:deltaL2}\\
L_{xy}^{(PQ)}(i\eta)-L_{xy}^{(PQ)}(0) & =\sum_{m}\int_{\text{BZ}}f_{m}\varepsilon_{\alpha\beta}\label{deltaL3}\nonumber \\
 & \qquad\times\partial_{\alpha}\!\left[E_{m}(\mathbf{k})A_{m,\beta}(\mathbf{k})\right]\\
L_{xy}^{(QQ)}(i\eta)-L_{xy}^{(QQ)}(0) & =\sum_{m}\int_{\text{BZ}}f_{m}\varepsilon_{\alpha\beta}\label{eq:deltaL4}\nonumber \\
 & \qquad\times\partial_{\alpha}\!\left[E_{m}^{2}(\mathbf{k})A_{m,\beta}(\mathbf{k})\right],
\end{align}
where $A_{m,\beta}(\mathbf{k})=\sum_{i}U_{mi}^{\dagger}(\mathbf{k})\partial_{\beta}U_{mi}(\mathbf{k})$
is the $\beta=x,y$ component of the Berry connection of the $m$-th
band, and $U_{m.i}$ is the matrix element of the unitary transformation
that relates the orbital basis to the energy basis, $|m\rangle=\sum_{i}U_{mi}|i\rangle$.
We introduce an integral over the density of states, $\int_{-\infty}^{\infty}\text{d}\epsilon\,\delta[\epsilon-E_{m}(\mathbf{k})+\mu]$
and change all energy dependence to the integration variable $\epsilon$,
\begin{align}
L_{xy}^{(PP)}(i\eta)-L_{xy}^{(PP)}(0) & =\int_{-\infty}^{\infty}d\epsilon\,f(\epsilon)\tilde{\sigma}_{xy}(\epsilon)\label{eq:LPP-1}\\
L_{xy}^{(PQ)}(i\eta)-L_{xy}^{(PQ)}(0) & =\int_{-\infty}^{\infty}d\epsilon\,f(\epsilon)[\epsilon\tilde{\sigma}_{xy}(\epsilon)+\tilde{\Sigma}_{xy}(\epsilon)]\label{eq:LPQ1}\\
L_{xy}^{(QQ)}(i\eta)-L_{xy}^{(QQ)}(0) & =\int_{-\infty}^{\infty}d\epsilon\,f(\epsilon)\nonumber \\
 & \qquad\times[\epsilon^{2}\tilde{\sigma}_{xy}(\epsilon)+2\epsilon\tilde{\Sigma}_{xy}(\epsilon)],\label{eq:LQQ1}
\end{align}
where 
\begin{align}
\tilde{\sigma}_{xy}(\epsilon) & =\int_{\text{BZ}}\sum_{m}\delta\left[\epsilon-E_{m}(\mathbf{k})+\mu\right]\varepsilon_{\alpha\beta}\partial_{\alpha}A_{m,\beta}(\mathbf{k})\\
\tilde{\Sigma}_{xy}(\epsilon) & =\int_{\text{BZ}}\sum_{m}\delta\left[\epsilon-E_{m}(\mathbf{k})+\mu\right]\varepsilon_{\alpha\beta}\label{eq:Sigma}\nonumber \\
 & \qquad\qquad\qquad\times[\partial_{\alpha}E_{m}(\mathbf{k})]A_{m,\beta}(\mathbf{k})
\end{align}
explicitly depend on the geometry of the $k$-space. 

We assume the energy dispersion of the bands $E_{m}(\mathbf{k})$
to be generic, with a finite velocity. Even though the term $\tilde{\Sigma}_{xy}(\epsilon)$
is proportional to the velocity of the quasiparticles and appears
to vanish in the flat band limit, this is not so \cite{Auerbach}.
To evaluate the integrand of Eq. (\ref{eq:Sigma}), we need to find
the contours of constant energy and change variables to coordinates
along the contour and perpendicular to it. In those variables, namely
$k_{\parallel}$ and $k_{\perp}$ for momenta parallel and perpendicular
to the energy contour respectively, $\text{d}E_{m}(\mathbf{k})=[\partial E_{m}(\mathbf{k})/\partial k_{\perp}]\text{d}k_{\perp}=v_{m}(\mathbf{k})\text{d}k_{\perp}$,
where $v_{m}$ is the velocity of the quasiparticles along the $\hat{k}_{\perp}$
direction. Hence, the element of volume in the momentum integrals
in Eq. (\ref{eq:Sigma}) can be recast as 
\begin{equation}
\text{d}^{2}k=\text{d}k_{\parallel}\text{d}k_{\perp}=\text{d}k_{\parallel}\text{d}E_{m}\,v_{m}^{-1}(\mathbf{k}),\label{eq:Id1}
\end{equation}
and scales inversely with the velocity of the quasiparticles $v_{m}(\mathbf{k})$.
At the same time, we note that the integrand in Eq. (\ref{eq:Sigma})
\begin{align}
\varepsilon_{\alpha\beta}[\partial_{\alpha}E_{m}(\mathbf{k})]A_{m,\beta}(\mathbf{k}) & =\left[v_{m}(\mathbf{k})\hat{\mathbf{k}}_{\perp}\times\mathbf{A}_{m}(\mathbf{k})\right]\cdot\hat{z}\nonumber \\
 & =v_{m}(\mathbf{k})\hat{\mathbf{k}}_{\parallel}\cdot\mathbf{A}_{m}(\mathbf{k})\label{eq:1d2}
\end{align}
is proportional to $v_{m}(\mathbf{k})$. The delta function in Eq.
(\ref{eq:Sigma}) constraints the integration along the constant energy
contour. Using Stokes theorem, 
\begin{align}
\tilde{\Sigma}_{xy}(\epsilon) & =\sum_{m}\oint_{E_{m}(\mathbf{k})-\mu=\epsilon}\frac{\text{d}k_{\parallel}}{(2\pi)^{2}}\hat{\mathbf{k}}_{\parallel}\cdot\mathbf{A}_{m}(\mathbf{k})\nonumber \\
 & =\sum_{m}\int_{E_{m}(\mathbf{k})-\mu\leq\epsilon}\frac{\text{d}k^{2}}{(2\pi)^{2}}\left[\partial_{\mathbf{k}}\times\mathbf{A}_{m}(\mathbf{k})\right]\cdot\hat{z}\nonumber \\
 & =\int_{-\infty}^{\epsilon}d\epsilon'\,\tilde{\sigma}_{xy}(\epsilon'),\label{eq:Sigma4}
\end{align}
where the last line implies the simple relationship $\tilde{\sigma}_{xy}(\epsilon)=\text{d}\tilde{\Sigma}_{xy}(\epsilon)/\text{d}\epsilon$.
The contribution $\tilde{\Sigma}_{xy}(\epsilon)$ remains finite when
we set $v_{m}(\mathbf{k})\to0$ in the flat band limit. Replacing
this result in Eq. (\ref{eq:LQQ1}) and integrating by parts, we have
the result
\begin{align}
L_{xy}^{(PQ)}(i\eta)-L_{xy}^{(PQ)}(0) & =-k_{B}T\label{eq:LPQ3}\\
 & \qquad\negthickspace\negthickspace\times\int_{-\infty}^{\infty}\text{d}x\,\frac{\text{d}f(x)}{\text{d}x}x\tilde{\Sigma}_{xy}(xk_{B}T)\\
L_{xy}^{(QQ)}(i\eta)-L_{xy}^{(QQ)}(0) & =-(k_{B}T)^{2}\label{eq:LQQ3}\nonumber \\
 & \qquad\negthickspace\negthickspace\times\int_{-\infty}^{\infty}\text{d}x\,\frac{\text{d}f(x)}{\text{d}x}x^{2}\tilde{\Sigma}_{xy}(xk_{B}T),
\end{align}
which allows us to extend the results of Ref. \cite{Bradlyn,Xiao-1}
to flat bands. The integration variable $x$ is dimensionless, $f(x)=(\text{e}^{x}+1)^{-1}$
and 
\begin{equation}
\tilde{\Sigma}_{xy}(xk_{B}T)=\int_{\text{BZ}}\sum_{m}\theta[xk_{B}T-E_{m}(\mathbf{k})+\mu]\Omega_{m}^{xy},\label{eq:Sigma3}
\end{equation}
with $\Omega_{m}^{xy}=\varepsilon_{\alpha\beta}\partial_{\alpha}A_{m,\beta}(\mathbf{k})$
the Berry curvature of band $m$, and $\theta$ is a step function.
The sum is restricted over the occupied bands in the zero temperature
limit, and thus $\tilde{\Sigma}_{xy}$ is proportional to the total
Chern number, $\lim_{T\to0}\tilde{\Sigma}_{xy}(xk_{B}T)=2\pi\mathcal{C}$.
In the opposite limit, $T\to\infty$, the sum over the bands is unrestricted
and gives the net Chern number of all bands in the BZ, which is zero,
$\lim_{T\to\infty}\tilde{\Sigma}_{xy}(xk_{B}T)=0$. The function $\tilde{\Sigma}_{xy}(xT)$
is therefore well behaved and generically describes the sum over the
Chern number of a selected number of bands in the flat band limit. 

The transverse part of the electric conductivity (\ref{eq:LPP}) can
be cast in the standard TKNN form \cite{Komissarov} (restoring $\hbar$),
\begin{equation}
\sigma_{xy}(0)=\frac{e^{2}}{\hbar}\int_{\text{BZ}}\sum_{m}f_{m}\Omega_{m}^{xy},\label{eq:sigmaxy}
\end{equation}
where $\mathcal{C}=2\pi\int_{\text{BZ}}\sum_{m}^{\text{occupied}}\Omega_{m}^{xy}$
gives the total Chern number of the bands. From the definition of
the conductivity tensors in Eq. (\ref{eq:alpha}) and (\ref{eq:kappa}),
the transverse part of the thermolelectric and thermal conductivities
is
\begin{equation}
\alpha_{xy}(0)=-\frac{k_{B}e}{\hbar}\!\int_{-\infty}^{\infty}\text{d}x\,\frac{\text{d}f(x)}{\text{d}x}x\tilde{\Sigma}_{xy}(xk_{B}T),\label{eq:alpha2}
\end{equation}
and
\begin{equation}
\kappa_{xy}(0)=-\frac{k_{B}^{2}T}{\hbar}\!\int_{-\infty}^{\infty}\text{d}x\,\frac{\text{d}f(x)}{\text{d}x}x^{2}\tilde{\Sigma}_{xy}(xk_{B}T).\label{eq:kappa2}
\end{equation}
Thanks to the magnetization subtraction, the integrals in Eq. (\ref{eq:alpha2})
and (\ref{eq:kappa2}) converge in the $T\to0$ limit, ensuring that
the thermal conductivity vanishes at zero temperature, as required
by the third law of thermodynamics. We note that due to the lack of
dispersion, the ground state entropy of flat bands scales with the
volume of the system, rather than being a constant. Nevertheless,
the rate of change of the entropy in the $T\to0$ limit still vanishes,
and so should all thermodynamic quantities defined by derivatives
of the entropy, such as the thermodynamic specific heat.

\subsection{Seebeck and Nernst coefficients}

Before we address the thermal and thermoelectric transport in flat
bands, we briefly review the general thermodynamic definition of particle
and heat currents that is applicable to systems with non-trivial quantum
geometry. Particle and heat currents are each created by two distinct
thermodynamic `forces', $X_{\gamma}^{P}$ and $X_{\gamma}^{Q}$, relating
to real space gradients in the local potentials and in temperature
\cite{Mahan}, 
\begin{align}
X_{\gamma}^{P} & =-\nabla_{\gamma}(\mu+V),\label{eq:XP}\\
X_{\gamma}^{Q} & =-\frac{1}{T}\nabla_{\gamma}T,\label{eq:XQ}
\end{align}
where $V$ is the bias voltage. The linear response of the particle/heat
currents to these forces is expressed in terms of the transport coefficients
$L_{\gamma\delta}^{(AB)}\equiv L_{\gamma\delta}^{(AB)}(\omega_{+})-L_{\gamma\delta}^{(AB)}(0),$
\begin{equation}
\left(\begin{array}{c}
J_{\gamma}^{P}\\
J_{\gamma}^{Q}
\end{array}\right)=\left(\begin{array}{cc}
L_{\gamma\delta}^{(PP)} & L_{\gamma\delta}^{(PQ)}\\
L_{\gamma\delta}^{(QP)} & L_{\gamma\delta}^{(QQ)}
\end{array}\right)\left(\begin{array}{c}
X_{\delta}^{P}\\
X_{\delta}^{Q}
\end{array}\right),\label{eq:onsager_p-1}
\end{equation}
where $J_{\gamma}^{P,Q}\equiv\langle j_{\gamma}^{P,Q}\rangle$ denotes
statistical average of current densities. Those coefficients obey
the Onsager reciprocity relation $L_{\gamma\delta}^{(PQ)}=L_{\gamma\delta}^{(QP)}$.
We assume that the chemical potential is uniformly distributed over
the material ($\nabla_{\gamma}\mu=0$). 

The heat transport properties are measured in actual experiments under
the condition of no particle current \cite{Behnia}. Enforcement of
this constraint permits expressing the electric potential gradient
in terms of the temperature gradient
\begin{equation}
J_{\gamma}^{P}=L_{\gamma\delta}^{(PP)}X_{\delta}^{P}+L_{\gamma\delta}^{(PQ)}X_{\delta}^{Q}=0.\label{eq:constraint}
\end{equation}
The heat current density under this constraint is given by
\begin{equation}
J_{\gamma}^{Q}=\mathcal{K}_{\gamma\delta}\nabla_{\delta}T,\label{eq:Jq2}
\end{equation}
with
\begin{equation}
\mathcal{K}_{\gamma\delta}=\frac{1}{T}[L_{\gamma\delta}^{(QQ)}+L_{\gamma\alpha}^{(PQ)}\mathcal{M}_{\alpha\delta}]\label{eq:Kappa}
\end{equation}
the thermal conductivity tensor and 
\begin{equation}
\mathcal{M}_{\gamma\delta}=-\frac{\varepsilon_{\gamma\alpha}L_{y\alpha}^{(PP)}L_{x\delta}^{(PQ)}-\varepsilon_{\gamma\alpha}L_{x\alpha}^{(PP)}L_{y\delta}^{(PQ)}}{L_{xx}^{(PP)}L_{yy}^{(PP)}-L_{xy}^{(PP)}L_{yx}^{(PP)}}.\label{eq:M}
\end{equation}
the thermoelectric tensor. Summation over $\alpha$ indices is implied. 

The thermoelectric properties are measured through the experimental
determination of the Seebeck and Nernst coefficients. The Seebeck
coefficient, also known as the thermopower, measures the ratio between
the voltage drop and a temperature gradient applied in the same direction,
which we arbitrarily choose to be $x$ \cite{Behnia}, $S=-\nabla_{x}V/\nabla_{x}T$,
in the condition where $J_{\gamma}^{P}=0$. Because of the zero particle
current constraint (\ref{eq:constraint}), the ratio between those
two gradients is expressed in terms of transport coefficients as
\begin{align}
S & =\frac{1}{eT}\mathcal{M}_{xx}\label{eq:S2}\nonumber \\
 & =-\frac{1}{eT}\frac{L_{yy}^{(PP)}L_{xx}^{(PQ)}-L_{xy}^{(PP)}L_{yx}^{(PQ)}}{L_{xx}^{(PP)}L_{yy}^{(PP)}-L_{xy}^{(PP)}L_{yx}^{(PP)}}.
\end{align}
In non-topological bands, where $L_{xy}^{(PP)}=0$, the Seebeck coefficient
reduces to the ratio between the thermoelectric and the electric longitudinal
conductivities defined in Eq. (\ref{eq:sigma}) and (\ref{eq:alpha}),
$S=\alpha_{xx}/\sigma_{xx}$. The latter is a well know expression
applicable to the case of metals in the low magnetic field limit and
also in trivial insulators \cite{Behnia}, but is not valid for Chern
insulators. The longitudinal transport coefficients of flat bands
are proportional to the broadening, $L_{\gamma\gamma}^{(AB)}\propto\eta$,
whereas the transverse coefficients $L_{xy}^{(AB)}$ are non-dissipative,
and hence broadening independent to leading order in $\eta$. It is
clear from Eq. (\ref{eq:S2}) that the Seebeck coefficient of flat
bands with non-zero Chern number is dominated by the transverse transport
coefficients, 
\begin{equation}
S=-\frac{1}{eT}\frac{L_{xy}^{(PQ)}}{L_{xy}^{(PP)}}+\mathcal{O}(\eta^{2}),\label{eq:S}
\end{equation}
and is topological up to non-universal corrections in second order
of the broadening $\eta$. 

In the same way, the Nernst coefficient measures the transverse voltage
drop produced by a thermal gradient,
\begin{align}
N & =-\frac{\nabla_{x}V}{\nabla_{y}T}=\frac{1}{eT}\mathcal{M}_{xy}.\label{eq:N-1}
\end{align}
In topologically trivial materials, it reduces to the ratio between
the transverse thermoelectric conductivity and the longitudinal conductivity,
$N=\alpha_{xy}/\sigma_{xx}$. The Nernst coefficient of flat Chern
bands is proportional to the broadening in leading order,
\begin{equation}
N=-\frac{L_{yy}^{(PP)}L_{xy}^{(PQ)}-L_{xy}^{(PP)}L_{yy}^{(PQ)}}{\left(L_{xy}^{(PP)}\right)^{2}}+\mathcal{O}(\eta^{2}).\label{eq:N2}
\end{equation}
In the next sections we will specifically consider two different families
of flat band Hamiltonians with non-trivial quantum geometry. 

\section{Two flat Chern bands}

The proceeding analysis is valid for any bands without a Fermi surface,
such as insulators. We now explicitly focus in the case of perfectly
flat bands, where quantum geometric effects are dominant. The simplest
example is the case of two flat Chern bands illustrated in Fig. 1a,
with the generic Hamiltonian 
\begin{equation}
\hat{h}(\mathbf{k})=\frac{\Delta}{2}\hat{\mathbf{d}}(\mathbf{k})\cdot\vec{\sigma},\label{eq:Ham}
\end{equation}
where $\hat{\mathbf{d}}(\mathbf{k})=(d_{1},d_{2},d_{3})$ is any three
dimensional unit vector function of $\mathbf{k}$ in a 2D BZ. The
eigenstates are labeled $|+,\mathbf{k}\rangle,|-,\mathbf{k}\rangle$
with energies
\begin{equation}
E_{\sigma}=\sigma\frac{\Delta}{2},\label{eq:Esigma}
\end{equation}
where $\sigma=\pm$. The Chern number of the bands is given by $\mathcal{C}_{\sigma}=2\pi\sigma\int_{\text{BZ}}\hat{\mathbf{d}}\cdot(\partial_{x}\hat{\mathbf{d}}\times\partial_{y}\hat{\mathbf{d}})$.
The DC particle transport coefficients of this system are written
as
\begin{equation}
L_{\gamma\delta}^{(PP)}=f_{-+}\left(\frac{4\eta}{\Delta}\int_{\text{BZ}}g_{-+}^{\gamma\gamma}-\int_{\text{BZ}}\Omega_{-+}^{\gamma\delta}\right).\label{eq:LPP2}
\end{equation}
All longitudinal DC transport coefficients are related by the relation
\begin{equation}
L_{\gamma\gamma}^{(QQ)}=-\mu L_{\gamma\gamma}^{(QP)}=\mu^{2}L_{\gamma\gamma}^{(PP)}\,,\label{eq:LQQ2}
\end{equation}
where $g_{-+}^{\gamma\delta}$ and $\Omega_{-+}^{\gamma\delta}$ are
the interband quantum metric and Berry curvature defined in Eq. (\ref{eq:gmn})
and (\ref{eq:Omegamn}) respectively. Defining the integral
\begin{equation}
\mathcal{F}^{(n)}(\mu,T)\equiv\int_{(-\frac{\Delta}{2}-\mu)/k_{B}T}^{(\frac{\Delta}{2}-\mu)/k_{B}T}\text{d}x\,\frac{\text{d}f(x)}{\text{d}x}x^{n},\label{eq:F}
\end{equation}
from Eq. (\ref{eq:LPQ3}) and (\ref{eq:LQQ3}), the other two remaining
transverse transport coefficients are (restoring $\hbar$)
\begin{equation}
L_{xy}^{(PQ)}=\frac{\mathcal{C}}{h}k_{B}T\mathcal{F}^{(1)}(\mu,T)\label{eq:LPQ3-1}
\end{equation}
and
\begin{equation}
L_{xy}^{(QQ)}=\frac{\mathcal{C}}{h}(k_{B}T)^{2}\mathcal{F}^{(2)}(\mu,T),\label{eq:LQQ3-1}
\end{equation}
where $\mathcal{C}=2\pi\int_{\text{BZ}}\Omega_{-}^{xy}$ is the Chern
number of the lower band. 

\subsection{Quantum metric lower bound}

It has been stated in specific analytic models \cite{Kruchkov,Mera,Ledwith}
that inequality (\ref{eq:g}) saturates the lower bound of the quantum
metric in the flat band limit. We can explicitly show that any generic,
single particle, two-level flat band model belonging to the unitary
symmetry class A of the classification table of topological insulators
\cite{Schneyder} saturates the lower bound of the quantum metric.
Hamiltonians in that class do not have any symmetries (time reversal,
chiral or particle-hole) and cover all Chern insulators in 2D. To
this end, we parametrize Hamiltonian (\ref{eq:Ham}) with a generic
meromorphic function $\chi(z)$, where $z\equiv k_{x}+ik_{y}$ \cite{Xie},
\begin{subequations}
\begin{align}
d_{\parallel}\equiv d_{1}+id_{2} & =\frac{2\chi(z)}{1+|\chi(z)|^{2}}\label{eq:dpar}\\
d_{\parallel}^{*}\equiv d_{1}-id_{2} & =\frac{2\chi^{*}(z)}{1+|\chi(z)|^{2}}\\
d_{3} & =\frac{1-|\chi(z)|^{2}}{1+|\chi(z)|^{2}},\label{eq:d3}
\end{align}
\end{subequations}and 
\begin{equation}
\partial_{x}=\partial_{z}+\partial_{\bar{z}},\hspace{0.5cm}\partial_{y}=i(\partial_{z}-\partial_{\bar{z}}).\label{eq:partials}
\end{equation}
This parameterization is general for systems of two flat-bands where
the Chern number for the lower band is positive. The eigenstates for
this system are given by
\begin{align}
|+\rangle & =\frac{1}{\sqrt{1+|\chi|^{2}}}\begin{pmatrix}1\\
\chi
\end{pmatrix},\\
|-\rangle & =\frac{1}{\sqrt{1+|\chi|^{2}}}\begin{pmatrix}\chi^{*}\\
-1
\end{pmatrix}.
\end{align}
Taking the partial derivatives with respect to $z$ and $\overline{z}$,
we can evaluate the matrix elements
\begin{align}
\langle+|\partial_{x}|-\rangle & =i\frac{\partial_{\bar{z}}\chi^{*}}{1+|\chi|^{2}}\\
\langle+|\partial_{y}|-\rangle & =\frac{\partial_{\bar{z}}\chi^{*}}{1+|\chi|^{2}},
\end{align}
and thus 
\begin{equation}
\Omega_{-}^{xy}=\frac{2|\partial_{z}\chi|^{2}}{(1+|\chi|^{2})^{2}},\hspace{0.2cm}\Omega_{+}^{xy}=-\frac{2|\partial_{z}\chi|^{2}}{(1+|\chi|^{2})^{2}}.
\end{equation}
Indeed, we also find 
\begin{equation}
g_{-+}^{xx}=g_{+-}^{xx}=g_{-+}^{yy}=g_{+-}^{yy}=\frac{|\partial_{z}\chi|^{2}}{(1+|\chi|^{2})^{2}}
\end{equation}
and conclude that $g_{\sigma}^{xx}+g_{\sigma}^{yy}=|\Omega_{\sigma}^{xy}|$,
$\sigma=\pm$. We note that $\Omega_{-}^{xy}>0$ in the whole BZ.
This property applies for instance to the flattened Haldane model
\cite{Haldane1988}, which is described by this parametrization. 

\begin{figure*}
\begin{centering}
\includegraphics[scale=0.5]{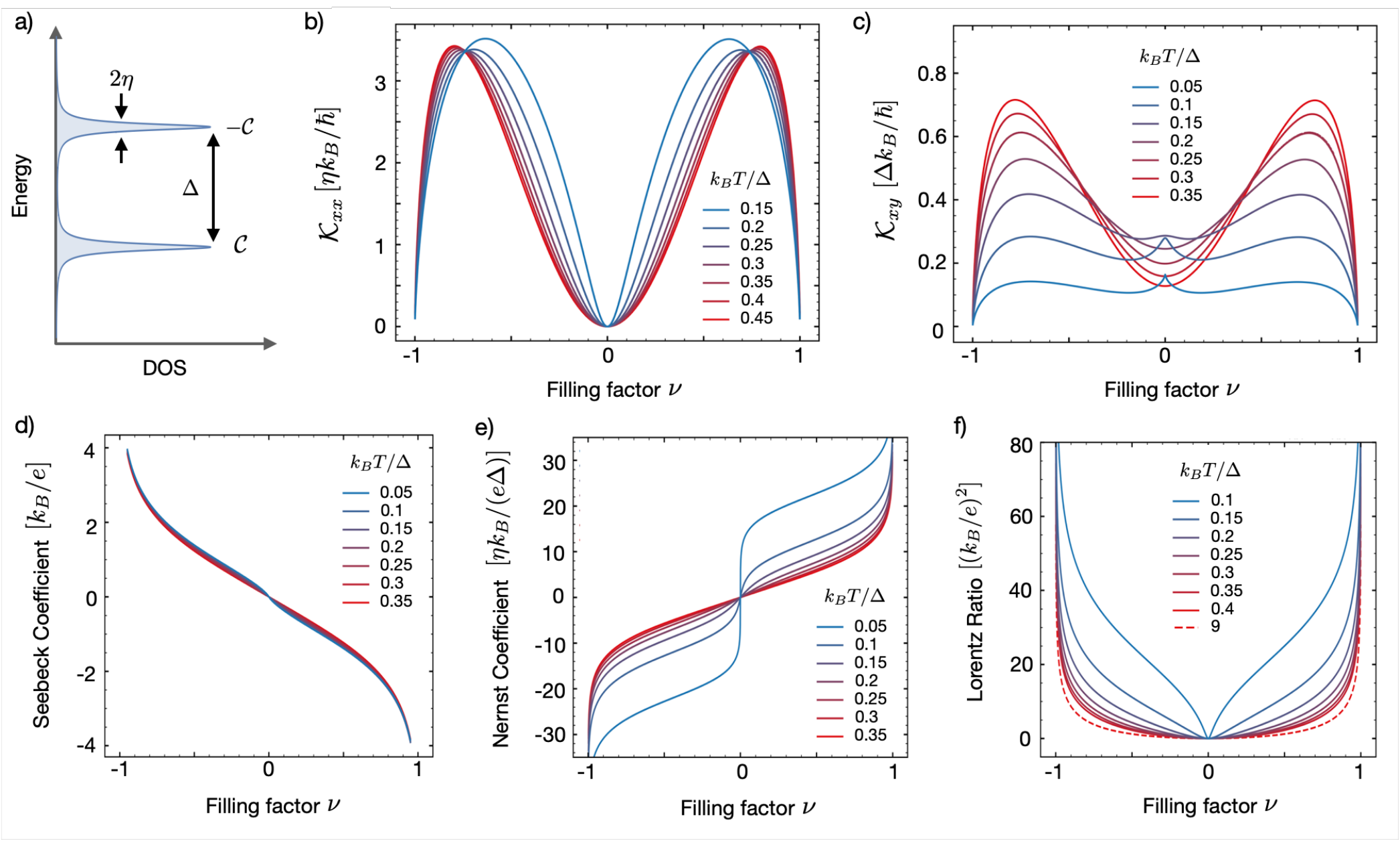}
\par\end{centering}
\caption{{\small Thermal and thermoelectric transport coefficients for two flat
Chern bands. a) Density of states of two flat bands with Chern numbers
$\pm\mathcal{C}$, level broadening $\eta$ and energy separation
$\Delta$. b) Longitudinal DC thermal conductivity as a function of
the combined filling factor of the bands, $\nu\in[-1,1]$, in units
of $\eta k_{B}/\hbar$. c) Transverse thermal conductivity in the
DC limit versus filling factor in units of $\Delta k_{B}/\hbar$.
d) Seedback coefficient (thermopower) $S$ and e) Nernst coefficient
$N$ in units of $k_{B}/e$ and $\eta k_{B}/(e\Delta)$ respectively,
versus filling factor $\nu$. f) Lorentz ratio $L\equiv\kappa_{\gamma\gamma}/(T\sigma_{\gamma\gamma}$)
versus filling factor. Different colors correspond to different temperature-gap
ratios ($k_{B}T/\Delta$), as indicated in the panels. $S$, $N$
and $L$ diverge logarithmically at integer fillings $\nu=-1,1$,
when the bands are either empty or completely filled, and vanish at
$\nu=0$ due to particle hole symmetry.}}
\end{figure*}

\subsection{Thermal and thermoelectric response of two flat Chern bands}

Using the previous results, we can now express equations (\ref{eq:LPP2})
and (\ref{eq:LQQ2}) in terms of the Chern number $\mathcal{C}$ of
the lower band, 
\begin{equation}
L_{\gamma\delta}^{(PP)}=f_{-+}\left(\frac{2\eta}{\Delta}\int_{\text{BZ}}|\Omega_{-}^{xy}|\delta_{\gamma\delta}-\mathcal{C}\varepsilon_{\gamma\delta}\right),\label{eq:PP2}
\end{equation}
where $f_{-+}=f_{-}-f_{+}$. Thus the longitudinal DC conductivity
of two generic flat Chern bands becomes (restoring $\hbar)$
\begin{equation}
\sigma_{\gamma\gamma}=\frac{e^{2}}{h}\frac{2\eta}{\Delta}f_{-+}|\mathcal{C}|.\label{eq:sigmaLong}
\end{equation}
The thermopower is
\begin{equation}
S=\frac{1}{eT}\mathcal{M}_{xx}=-\frac{2k_{B}}{ef_{-+}}\mathcal{F}^{(1)}(\mu,T),\label{eq:S4}
\end{equation}
and the Nernst coefficient has the form
\begin{align}
N & =\frac{1}{eT}\mathcal{M}_{xy}\nonumber \\
 & =\frac{1}{eT}\frac{\eta}{\Delta}\text{sign}(\mathcal{C})\left(2\mu+\frac{4T}{f_{-+}}\mathcal{F}^{(1)}(\mu,T)\right).\label{eq:N3}
\end{align}
Finally, the longitudinal components of the DC thermal conductivity
tensor are given by 
\begin{align}
\mathcal{K}_{_{\gamma\gamma}} & =\frac{1}{\hbar T}\left[L_{\gamma\gamma}^{(QQ)}+L_{\gamma x}^{(PQ)}M_{x\gamma}+L_{\gamma y}^{(PQ)}M_{y\gamma}\right]\nonumber \\
 & =\frac{2k_{B}\eta}{\hbar\Delta}|\mathcal{C}|\left\{ \frac{\mu^{2}}{k_{B}T}f_{-+}-\frac{k_{B}T}{f_{-+}}\left[\mathcal{F}^{(1)}(\mu,T)\right]^{2}\right\} ,\label{eq:Kappaxx2}
\end{align}
whereas the transverse component is 
\begin{align}
\mathcal{K}_{xy} & =\frac{1}{\hbar T}\left[L_{xy}^{(QQ)}+L_{xx}^{(PQ)}M_{xy}+L_{xy}^{(PQ)}M_{yy}\right]\nonumber \\
 & =\frac{k_{B}^{2}T}{\hbar}\mathcal{C}\left\{ \frac{\left[\mathcal{F}^{(1)}(\mu,T)\right]^{2}}{f_{-+}}-\mathcal{F}^{(2)}(\mu,T)\right\} +\mathcal{O}(\eta^{2}).\label{eq:Kappaxy2}
\end{align}

We illustrate the broadened density of states for two flat Chern bands
separated by an energy gap $\Delta$ in Fig. 1a. The two flat bands
satisfy the relation $f_{+}+f_{-}=\nu+1$, with $\nu\in[-1,1]$ the
filling factor of the bands. Inverting this relation gives $\mu(\nu,T)$
as a function of filling and temperature. In Fig. 1b we show the longitudinal
DC thermal conductivity $\mathcal{K}_{\gamma\gamma}$ versus the filling
factor for two flat bands with Chern numbers $\pm\mathcal{C}$. $\mathcal{K}_{\gamma\gamma}$
is exactly zero for all temperatures at integer filling factors ($\nu=-1,0,1)$.
At $\nu=-1,1$, the bands are either empty or full and hence have
no transport. At $\nu=0$ the longitudinal thermal current vanishes
due to particle-hole symmetry, as a result of the quasiparticles carrying
zero energy \cite{Yang}. Thermal conductivity is maximal at partial
filling factors of the flat bands. The different temperatures are
shown in different colors assuming constant broadening $\eta$, which
is treated as a free parameter. At low temperatures, $k_{B}T\lesssim\eta$,
the broadening is expected to vanish with temperature as the life-time
of the quasiparticles grows. The longitudinal components of the thermoelectric
and thermal conductivities are expected on physical grounds to vanish
in the same limit. 

The transverse component of the thermal conductivity tensor is shown
in Fig. 1c as a function of the total filling factor of two flat Chern
bands for different temperatures. The Seebeck (Fig. 1d) and Nernst
coefficients (Fig. 1e) show logarithmic divergences at $\nu=-1,1$,
when the bands are either completely empty or filled, and are zero
at half filling ($\nu=0$), when the bands have particle-hole symmetry.
The Seedback coefficient for two flat bands has a very weak temperature
dependence, illustrated by the near collapsation of the curves drawn
in different colors into a single curve. The Nernst coefficient is
proportional to $\eta/T$. Its temperature scaling at low temperature
depends on the microscopic details of the inelastic scattering mechanisms
in the solid. 

Finally we comment on the violation of the Wiedmann-Franz law for
two flat bands. The Lorentz ratio $L$ is defined as the ratio between
the longitudinal Onsager thermal conductivity $\kappa_{\gamma\gamma}$
and the longitudinal electric conductivity $\sigma_{\gamma\gamma}$
times temperature, 
\begin{equation}
L\equiv\frac{\kappa_{\gamma\gamma}}{T\sigma_{\gamma\gamma}}=\frac{L_{\gamma\gamma}^{(QQ)}}{e^{2}T^{2}L_{\gamma\gamma}^{(PP)}}=\frac{\mu^{2}(\nu,T)}{e^{2}T^{2}}.\label{eq:WFL}
\end{equation}
In the low temperature limit $k_{B}T\ll\Delta$ one can approximate
$\mu(\nu,T)\approx\text{sign}(\nu)\left[\Delta/2+k_{B}T\log\left(|\nu|/(1-|\nu|)\right)\right]$
for non-zero $\nu$, $(\nu\in[-1,1])$. At half-filling ($\nu=0$)
the Fermi level sits half-way between the bands and the chemical potential
$\mu$ is zero. The Lorentz ratio diverges in the low temperature
limit $\lim_{T\to0}L\propto1/T^{2}$ away from half-filling. This
is in strong violation of the Wiedmann-Franz law for metals, where
the Lorentz ratio is $\frac{\pi^{2}}{3}\left(\frac{k_{B}}{e}\right)^{2}$.
$L$ also diverges logarithmically at integer fillings $\nu=\pm1$,
and is zero at half filling for any temperature, as shown in Fig.
1f. 

\section{Generalized flattened Lieb model}

Let us now examine the results of sec. II in the three band case.
We consider an extension of the Lieb model \cite{Lieb}, which is
a simple topological Hamiltonian for three bands. The Lieb model has
two Chern bands with opposite Chern numbers, each described by massive
Dirac fermions in the continuum limit, and one perfectly flat band
with zero Chern number half-way in between. 

The flattened version of the Lieb model can be generalized to represent
any three equally spaced topological flat bands with zero total Chern
number, where the middle band is topologically trivial,
\begin{equation}
\hat{h}(\mathbf{k})=\Delta\left(\begin{array}{ccc}
0 & d_{1}(\mathbf{k}) & d_{2}(\mathbf{k})\\
d_{1}(\mathbf{k}) & 0 & id_{3}(\mathbf{k})\\
d_{2}(\mathbf{k}) & -id_{3}(\mathbf{k}) & 0
\end{array}\right),\label{eq:Lieb}
\end{equation}
with $\hat{\mathbf{d}}(\mathbf{k})=(d_{1},d_{2},d_{3})$ describing
an arbitrary 3D unit vector of functions of momentum $\mathbf{k}$.
The energy spectrum of Hamiltonian (\ref{eq:Lieb}) is
\begin{equation}
E_{m}=m\Delta,\label{eq:E3}
\end{equation}
with $m=0,\pm1$. As in the two-level flat band model, this family
of Hamiltonians also belongs to the topological class A in the periodic
table, which includes all Chern insulators. 

\subsection{Quantum metric }

Labeling the corresponding eigenkets as $|m\rangle$, with the periodic
part of the corresponding Bloch eigenfunctions $\langle\mathbf{r}|m\rangle\equiv\langle\mathbf{r}|u_{m},\mathbf{k}\rangle$,
we can resort to the same parametrization (\ref{eq:dpar})$-$(\ref{eq:d3})
we used in the two band model. The eigenkets of (\ref{eq:Lieb}) are
\begin{align}
|+\rangle & =\frac{1}{\sqrt{2}(1+|\chi|^{2})}\left(\begin{array}{c}
-2i\chi\\
-i(1+\chi^{2})\\
1-\chi^{2}
\end{array}\right)\label{eq:+}\\
|0\rangle & =\frac{1}{1+|\chi|^{2}}\left(\begin{array}{c}
i(1-|\chi|^{2})\\
i(\chi-\chi^{*})\\
\chi+\chi^{*}
\end{array}\right)\label{eq:0}\\
|-\rangle & =\frac{1}{\sqrt{2}(1+|\chi|^{2})}\left(\begin{array}{c}
-2i\chi^{*}\\
i(1+\chi^{*2})\\
1-\chi^{*2}
\end{array}\right),\label{eq:-}
\end{align}
with $\chi(z)$ a meromorphic function of $z=k_{x}+ik_{y}$. Calculating
the matrix elements
\begin{align}
\langle-|\partial_{z}|0\rangle & =\frac{\sqrt{2}\partial_{z}\chi}{1+|\chi|^{2}}\\
\langle+|\partial_{\overline{z}}|0\rangle & =\frac{\sqrt{2}\partial_{\overline{z}}\chi^{*}}{1+|\chi|^{2}},
\end{align}
and $\langle-|\partial_{\overline{z}}|0\rangle=\langle-|\partial_{z}|+\rangle=\langle-|\partial_{\overline{z}}|+\rangle=\langle+|\partial_{z}|0\rangle=0$,
we find the interband Berry curvatures,
\begin{equation}
\Omega_{-,0}^{xy}=-\Omega_{+,0}^{xy}=\frac{4|\partial_{z}\chi|^{2}}{(1+|\chi|^{2})^{2}},\qquad\Omega_{-,+}^{xy}=0.\label{eq:Omegaxy3}
\end{equation}
The quantum metric is also the magnitude of $\Omega$,
\begin{align}
g_{-,0}^{xx} & =g_{+,0}^{xx}=g_{-,0}^{yy}=g_{+,0}^{yy}=\frac{2|\partial_{z}\chi|^{2}}{(1+|\chi|^{2})^{2}},\label{eq:g2}
\end{align}
and $g_{-,+}^{xx}=g_{-,+}^{yy}=0$. 

It is clear from Eq. (\ref{eq:Omegaxy3}) and (\ref{eq:g2}) that
the lower bound of the interband quantum metric is saturated, $g_{m0}^{xx}+g_{m0}^{yy}=|\Omega_{m0}^{xy}|$,
with $m=\pm$. Since $g_{m}^{\gamma\gamma}=\sum_{n\neq m}g_{mn}^{\gamma\gamma}$,
$\Omega_{m}^{xy}=\sum_{n\neq m}\Omega_{mn}^{xy}$ and $g_{-+}^{\gamma\gamma}=\text{\ensuremath{\Omega}}_{-+}^{\gamma\gamma}=0$,
it immediately follows that the bound is also saturated for the trace
of the quantum metric of bands $m=\pm$, although not for $m=0$,
since
\begin{equation}
\text{tr}g_{0}^{\gamma\gamma}=\frac{8|\partial_{z}\chi|^{2}}{(1+|\chi|^{2})^{2}}>|\Omega_{0}^{xy}|=0.\label{eq:trace2}
\end{equation}

\subsubsection{$2S+1$ flat bands}

We now comment on the saturation of the quantum metric bound for an
arbitrary number of flat Chern bands. The previous results can be
extended to $2S+1$ flat bands,
\begin{equation}
\hat{h}(\mathbf{k})=\Delta\hat{\mathbf{n}}(\mathbf{k})\cdot\mathbf{S},\label{eq:h4}
\end{equation}
where $\hat{\mathbf{n}}(\mathbf{k})=(n_{1},n_{2},n_{3})$ is a 3D
unit vector field in a 2D BZ and \textbf{$\mathbf{S}=(S_{1},S_{2},S_{3})$
}are generators of SU(2) written as $(2S+1)\times(2S+1)$ matrices,
where $\mathbf{S}^{2}=S(S+1)\mathds{1}$. The eigenspectrum are $2S+1$
equally spaced flat bands with energy
\begin{equation}
E_{m}=m\Delta,\qquad m\in-S,-S+1,\ldots S,\label{eq:E}
\end{equation}
with $S$ integer of half-odd integer. The wavefunction of band $m$
is given by a spin coherent state
\begin{equation}
|\hat{\mathbf{n}},m\rangle\equiv R(\hat{\mathbf{n}})|S,m\rangle,\label{eq:coherent}
\end{equation}
where $S_{3}|S,m\rangle=m|S,m\rangle$. The unitary spin rotation
operator which rotates $S_{3}$ to $\hat{\mathbf{n}}\cdot\mathbf{S}$
is $R(\hat{\mathbf{n}})=\text{e}^{i\phi S_{3}}\text{e}^{i\theta S_{2}}$,
where $(\phi,\theta)$ are the polar coordinates of the unit vector
$\hat{\mathbf{n}}$. 

The Berry curvature of band $m$ is (see Appendix B)
\begin{equation}
\Omega_{m}=2m(\partial_{x}\hat{\mathbf{n}}\times\partial_{y}\hat{\mathbf{n}})\cdot\hat{\mathbf{n}}.\label{eq:Omega}
\end{equation}
The Chern number of band $m$ is hence
\begin{equation}
\mathcal{C}_{m}=2\pi\int_{\text{BZ}}\Omega_{m}=2mQ,\label{eq:C3}
\end{equation}
where $Q=2\pi\int_{\text{BZ}}(\partial_{x}\hat{\mathbf{n}}^{Q}\times\partial_{y}\hat{\mathbf{n}}^{Q})\cdot\hat{\mathbf{n}}^{Q}$
is the Pontryagin integer mapping $\hat{\mathbf{n}}^{Q}$ of the BZ
torus to the unit sphere, i.e. the total skyrmion number. For $S$
integer, the spectrum has an odd number of bands ($2S+1$), with the
middle band ($m=0$) being topologically trivial. The spectrum has
an even number of bands for half-odd integer $S$, all of them with
a non-zero Chern number for a given non-zero $Q$. This model thus
describes either $2S$ or $2S+1$ flat Chern bands for $S$ integer
or half-odd integer, respectively.

As shown in Appendix B, the quantum metric tensor of band $m$ is
\begin{equation}
g_{m}^{\gamma\delta}=\frac{S(S+1)-m^{2}}{2}\left(\partial_{\gamma}\hat{\mathbf{n}}\right)\cdot\left(\partial_{\delta}\hat{\mathbf{n}}\right).\label{eq:g6}
\end{equation}
The trace of the quantum metric is 
\begin{align}
\text{tr}g_{m}^{\gamma\delta} & =\frac{S(S+1)-m^{2}}{2}\sum_{\gamma}(\partial_{\gamma}\hat{\mathbf{n}})^{2}\nonumber \\
 & =\frac{S(S+1)-m^{2}}{m}|\Omega_{m}|\nonumber \\
 & +\frac{S(S+1)-m^{2}}{2}\sum_{\gamma}\left(\partial_{\gamma}\hat{\mathbf{n}}+\sum_{\delta}\varepsilon_{\gamma\delta}\hat{\mathbf{n}}\times\partial_{\delta}\hat{\mathbf{n}}\right)^{2},\label{eq:trace}
\end{align}
where the second equality follows from a stablished identity for the
non-linear sigma model \cite{Polyakov}. 

\begin{figure*}
\begin{centering}
\includegraphics[scale=0.5]{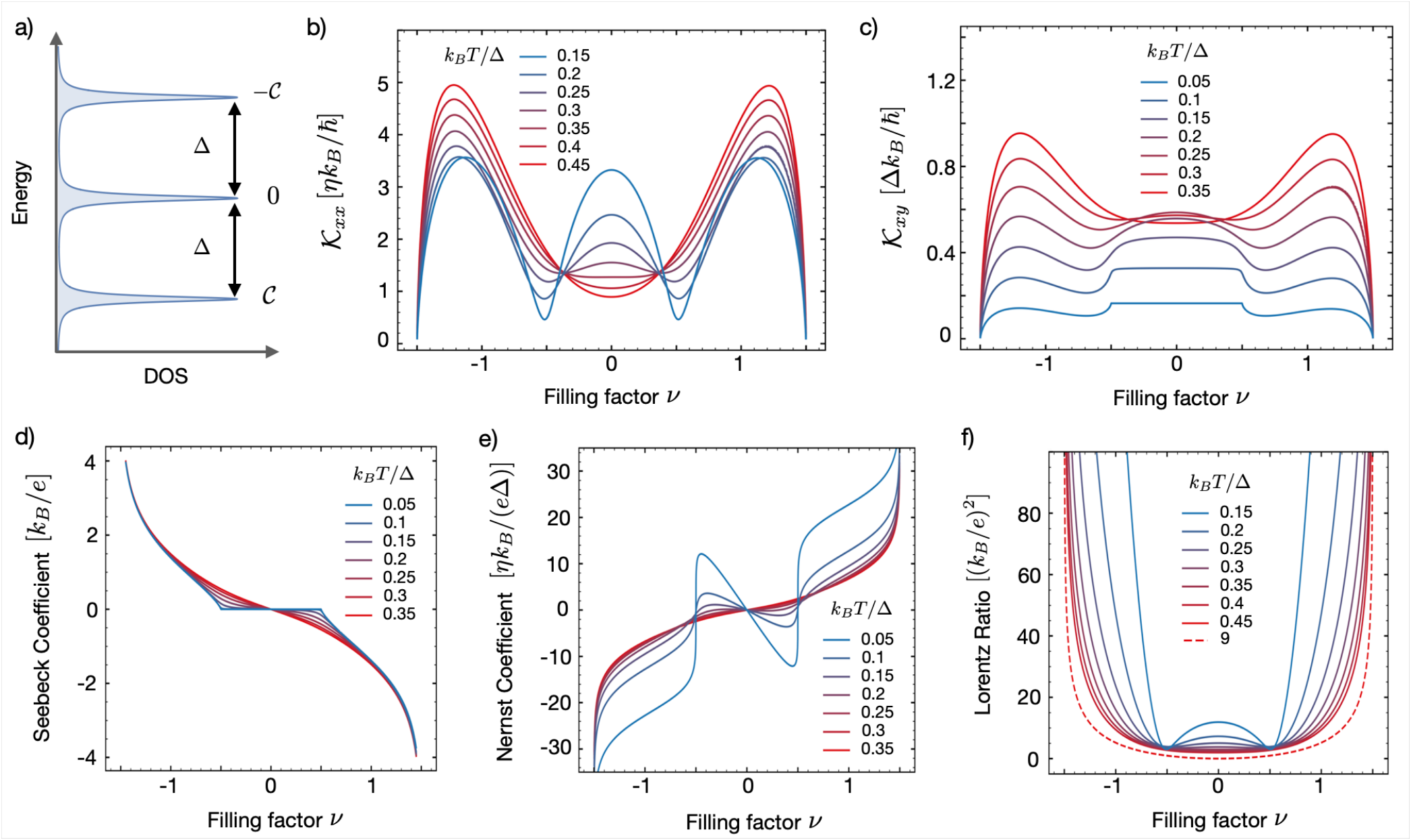}
\par\end{centering}
\caption{{\small a) Density of states for the generalized flattened Lieb model,
with three equally spaced flat bands. The extremal bands at the bottom
and at the top of the spectrum are topological, with Chern numbers
$\pm\mathcal{C}$, and the middle one is topologically trivial. b)
Longitudinal ($\mathcal{K}_{xx}$) and c) transverse ($\mathcal{K}_{xy}$)
thermal conductivity times temperature $T$, in units of $\eta k_{B}/\hbar$
and $\Delta k_{B}/\hbar$, respectively, versus filling factor $\nu\in[-\frac{3}{2},\frac{3}{2}]$.
d) Seebeck coefficient (thermopower) and e) Nernst coefficient, in
units of $\eta k_{B}/(e\Delta)$, versus filling factor $\nu$. The
zero temperature plateaus for $\mathcal{K}_{xy}$ and the Seebeck
coefficient $S\propto L_{xy}^{(PQ)}/L_{xy}^{(PP)}$ for $\nu\in[-\frac{1}{2},\frac{1}{2}]$
reflect the topologically trivial nature of the middle band (see text).
f) Lorentz ratio $L\equiv\kappa_{\gamma\gamma}/(T\sigma_{\gamma\gamma})$
versus $\nu$. Different colors correspond to different temperature-gap
ratios. }}
\end{figure*}

The local quantum geometry saturation condition $\text{tr}g_{m}^{\gamma\delta}=|\Omega_{m}|$
for Hamiltonian (\ref{eq:h4}) can be satisfied under two conditions:
\begin{enumerate}
\item That we consider the two extremal bands, $m=-S,S$, where the Chern
number is quantized in units of $2S$, $\mathcal{C}_{\pm S}=\pm2SQ=\text{integer}.$
\item The Hamiltonian function $\hat{\mathbf{n}}(\mathbf{k})$ must be a
minimal energy configuration of the non-linear sigma model for any
Pontryagin index $Q$, which will annihilate the second term in Eq.
(\ref{eq:trace}) by satisfying the differential equation,
\begin{equation}
\partial_{\gamma}\hat{\mathbf{n}}=-\sum_{\delta}\varepsilon_{\gamma\delta}\hat{\mathbf{n}}\times\partial_{\delta}\hat{\mathbf{n}}.\label{eq:diff}
\end{equation}
\end{enumerate}
Using the parametrization (\ref{eq:dpar})$-$(\ref{eq:d3}) through
the function $\chi(z)$ and the complex variable $z=k_{x}+ik_{y}$,
$\bar{z}=z^{*}$, then Eq. (\ref{eq:diff}) can be written as,
\begin{equation}
\partial_{\bar{z}}\chi(z)=0,\label{eq:diff2}
\end{equation}
which implies that $\chi$ is a meromorphic function of $z$. On the
infinite BZ, with $\chi\to1$ at $\lim z\to\infty$, this equation
is solved for any Pontryagin index $Q$ defined by any set of zeros
and poles, $\chi^{Q}=\prod_{i=1}^{Q}(z-a_{i})/(z-b_{i})$. \textcolor{black}{In
conclusion, in the family of flat band Hamiltonians describing SU(2)
coherent states considered in this paper, the saturation condition
is restricted to extremal flat bands $m=\pm S$, with Chern numbers
$\mathcal{C}=\pm2SQ$, whose Hamiltonian Eq. (\ref{eq:h4}) is defined
by a meromorphic function in the reciprocal space. }

\textcolor{black}{We note that a meromorphic function has no essential
singularities and its poles are isolated. Continuously moving the
poles away from the edge of the BZ, one can always construct a contour
around the BZ that skips all poles. On a torus, which can be mapped
into a rectangular strip with periodic boundary conditions, any doubly
periodic function will have a zero contour integral around the edge,
since opposite edges will have the same values of the function but
will be traversed by the contour in opposite directions. Therefore,
if $\hat{\mathbf{n}}(\mathbf{k})$ is periodic in the BZ and meromorphic
it cannot have a single simple pole, and thus cannot have Chern number
$\mathcal{C}=\pm1$ \cite{Jian}. }

\textcolor{black}{The argument above appears to exclude Hamiltonians
with Chern numbers $\mathcal{C}=\pm1$ in the two band case ($S=\frac{1}{2})$
as possible candidates for the saturation of the lower bound. We point
out that this is not the case, since Chern insulator Hamiltonians
do not need to be periodic in the reciprocal lattice. For instance,
the flattened Haldane model is periodic in the reciprocal torus of
area 3 times the BZ. This model allows for a Chern number 3, which
translates to a Hall coefficient $\mathcal{C}=1$ for the band in
the BZ. Our conclusions regarding the saturation of the quantum metric
lower bound thus apply for flat bands with any integer Chern number. }

\subsection{Thermal and thermoelectric responses of the generalized flattened
Lieb model}

Replacement of the above results in equations (\ref{eq:LPP}$-$\ref{eq:LQQ}),
(\ref{eq:LPP-1}) and (\ref{eq:LPQ3}$-$\ref{eq:Sigma3}) gives the
DC transport coefficients for the generalized flattened Lieb model.
The particle transport coefficients have the same form as in Eq. (\ref{eq:PP2}),
resulting in the same conductivity tensor. Defining the integral
\begin{equation}
\mathcal{G}^{(n)}(\mu,T)=\int_{(-\Delta-\mu)/k_{B}T}^{(\Delta-\mu)/k_{B}T}\text{d}x\,\frac{\text{d}f(x)}{\text{d}x}x^{n},\label{eq:F3}
\end{equation}
the Seebeck and Nernst coefficients are 
\begin{equation}
S=\frac{1}{eT}M_{xx}=-\frac{k_{B}}{e}\frac{1}{f_{-+}}\mathcal{G}^{(1)}(\mu,T),\label{eq:S4-1}
\end{equation}
and
\begin{align}
N & =\frac{1}{eT}M_{xy}\nonumber \\
 & =\frac{\eta}{\Delta}\frac{1}{eT}\text{sign}(\mathcal{C})\left[k_{B}T\frac{2}{f_{-+}}\mathcal{G}^{(1)}-\Delta\frac{f_{0+}+f_{0-}}{f_{-+}}+2\mu\right],\label{eq:N5}
\end{align}
respectively. The longitudinal thermal conductivity $\mathcal{K}_{\gamma\gamma}$
follows directly from Eq. (\ref{eq:Kappa}) with the substitution
of the the thermoelectric coefficients $M_{xx},$ $M_{xy}$, given
explicitly in Eq. (\ref{eq:S4-1}) and (\ref{eq:N5}), and of the
transport coefficients 
\begin{equation}
L_{\gamma\gamma}^{(PQ)}=\eta|\mathcal{C}|\left[(f_{0+}+f_{0-})-\frac{2\mu}{\Delta}f_{-+}\right],\label{eq:L6}
\end{equation}
and
\begin{equation}
L_{\gamma\gamma}^{(QQ)}=\frac{\eta}{2}\Delta|\mathcal{C}|\left[\frac{4\mu}{\Delta}(f_{-0}+f_{+0})+\left(1+\frac{4\mu^{2}}{\Delta^{2}}\right)f_{-+}\right].\label{eq:LQQ6}
\end{equation}
The transverse thermal conductivity can be explicitly expressed in
a compact form, 
\begin{equation}
\mathcal{K}_{xy}=\frac{1}{\hbar}k_{B}^{2}T\mathcal{C}\left(-2\mathcal{G}^{(2)}(\mu,T)+4\frac{[\mathcal{G}^{(1)}(\mu,T)]^{2}}{f_{-+}}\right).\label{eq:Kappa6}
\end{equation}

The thermal conductivity for the generalized flattened Lieb model
is shown in Fig. 2. The three bands can hold at most three electrons
and satisfy $\sum_{\sigma=-1}^{1}f_{\sigma}=\nu+\frac{3}{2}$, with
$\nu\in[-\frac{3}{2},\frac{3}{2}]$. In panel 2b we show the longitudinal
thermal conductivity $\mathcal{K}_{xx}$ versus filling factor of
the bands $\nu$ for different temperatures, which are indicated by
different colors. $\mathcal{K}_{xx}$ is suppressed at fillings $\nu=\pm\frac{1}{2}$
in the low temperature regime $k_{B}T\ll\Delta$, when the chemical
potential sits in between two levels and the system is incompressible.
For $\nu\in(-\frac{1}{2},\frac{1}{2})$ the system is metallic and
thermal transport is parametrically enhanced in the low temperature
limit compared to other filling factors outside of that range. 

The transverse thermal conductivity $\mathcal{K}_{xy}$ (see Fig.
2c) and the Seebeck coefficient $S\propto L_{xy}^{(PQ)}/L_{xy}^{(PP)}$
(Fig. 2d) have plateaus as a function of the filling factor in the
low temperature regime for $\nu\in[-\frac{1}{2},\frac{1}{2}]$, when
the topologically trivial band is occupied. In that range of filling
factors, $S=0$ in the $T\to0$ limit, whereas the Nernst coefficient
$N$ has characteristic oscillations as a function of the filling
factor around $\nu=0$. Finally, the Lorentz ratio ratio between the
Onsager thermal conductivity and the charge conductivity, $L=\kappa_{\gamma\gamma}/(T\sigma_{\gamma\gamma})=L^{(QQ)}/[e^{2}T^{2}L_{\gamma\gamma}^{(PP)}]$,
diverges in the low temperature limit as $1/T^{2}$ at all filling
factors, except for $\nu=\pm\frac{1}{2}$ (see Fig. 2f), which are
incompressible states with thermally activated transport. At very
large temperatures (dashed line in Fig. 2f), the Lorentz ratio assymptotically
approaches zero at $\nu=0$. At any temperature, $L$ diverges logarithmically
near the minimum and maximum occupation of the bands ($\nu=\pm\frac{3}{2})$,
similarly to the two band case.

\section{Discussion}

We have addressed the thermal and thermoelectric transport in the
clean limit for families of flat Chern band Hamiltonians. Our longitudinal
DC heat transport responses assume the presence of inelastic scattering,
which permits the system to thermalize. Our charge conductivity tensor
reproduces previous results in the literature \cite{Huhtinen,Mitscherling,Mera,Mitscherling2},
which found a finite DC longitudinal conductivity for flat bands in
the zero temperature limit that is proportional to the broadening
and to the integral of the quantum metric in the BZ. We also calculated
the thermal conductivity tensor, and the Nernst and Seebeck coefficients
as a function of the filling and temperature, clarifying the importance
of the magnetization subtraction for the transverse part of the thermal
and thermoelectric transport coefficients. We show that while the
longitudinal part of the DC transport coefficients is dominated by
interband processes in the flat band limit, the transverse part is
on-shell. That requires taking a proper order of limits, in which
the magnetization subtraction is taken before the flat band limit. 

Thermal and thermoelectric currents have been previously calculated
in Ref. \cite{Kruchkov} for two disordered flat Chern bands at quarter
filling, i.e. when the lower band is half-filled. Besides addressing
the problem in the clean limit, our results differ from those in two
important ways. The first one is that we transparently calculated
the thermal and thermoelectric responses using the Lehman representation,
which as we show in Appendix C, leads to results that are fully consistent
with the ones calculated with the Green's function formalism in the
clean limit. The earlier results were calculated using the Kubo-Streda
formula, which gives results that conflict \cite{Huhtinen} with the
temperature dependence of the DC longitudinal charge conductivity
found in this work and in previous calculations \cite{Huhtinen,Mitscherling,Mera,Mitscherling2}.
The second difference is that our results describe the condition of
zero particle flow, which correspond to the experimental situation.
As we pointed out in Sec. II C, the standard formulas \cite{Behnia}
conventionally used for calculating the thermal conductivity, the
Seebeck and the Nernst coefficients in metals at weak magnetic fields
and insulators are not applicable to Chern insulators. 

We also examined the saturation of the local bound of the quantum
metric for a general family of Hamiltonians describing SU(2) coherent
states, which produce an arbitrary number of equally spaced flat bands.
We showed that in this family saturation of the bound can be found
only in the extremal bands, at the very bottom and top of the energy
spectrum, under the condition that the momentum dependence of the
Hamiltonian is described by a meromorphic function in the BZ. 

We acknowledge C. Bolech, T. Holder, I. Soddermann, J. Moore \textcolor{black}{and
D. Arovas} for illuminating discussions. BU was supported by NSF grant
No. DMR-2024864. BU thanks the Aspen Center for Physics, where this
work was partially completed. \textcolor{black}{This paper is dedicated
to the memory of Assa Auerbach. }

\appendix

\section{DC limit}

The $\omega\to0$ limit of Eq. (\ref{eq:L1}) needs to be taken with
some care. The summand becomes antisymmetric in that limit, yet the
$1/\omega$ coefficient gives a finite result. If $s_{mn}$ is some
arbitrary symmetric matrix element, then \begin{widetext}
\begin{align*}
\lim_{\omega\rightarrow0}\sum_{mn}\frac{f_{mn}s_{mn}}{\omega[(\omega_{mn}+\omega)^{2}+\eta^{2}]} & =\lim_{\omega\rightarrow0}\sum_{mn}f_{m}\left(\frac{s_{mn}}{\omega[(\omega_{mn}+\omega)^{2}+\eta^{2}]}-\frac{s_{mn}}{\omega[(\omega_{mn}-\omega)^{2}+\eta^{2}]}\right)\\
 & =\sum_{mn}s_{mn}f_{m}\lim_{\omega\rightarrow0}\frac{(\omega-\omega_{mn})^{2}-(\omega+\omega_{mn})^{2}}{\omega[(\omega+\omega_{mn})^{2}+\eta^{2}][(\omega-\omega_{mn})^{2}+\eta^{2}]}\\
 & =-4\sum_{mn}f_{m}\frac{\omega_{mn}s_{mn}}{(\omega_{nm}^{2}+\eta^{2})^{2}}.
\end{align*}
\end{widetext}Substitution of $s_{mn}\to\text{Re}[j_{\gamma,mn}^{A}j_{\delta,mn}^{B}]$
or $s_{mn}\to\omega_{mn}\text{Im}[j_{\gamma,mn}^{A}j_{\delta,mn}^{B}]$
in Eq. (\ref{eq:L1}) followed by the leading order expansion in $\eta$
gives Eq. (\ref{eq:ReL2-1}).

\section{Saturation of the quantum metric bound}

The Berry curvature of band $m$ in the generic SU(2) symmetric model
of Hamiltonian (\ref{eq:h4}) is
\begin{equation}
\Omega_{m}(\mathbf{k})=2\text{Im}\langle S,m|(\partial_{x}R^{\dagger})(\partial_{y}R)|S,m\rangle-(x\rightarrow y),\label{eq:Omega3}
\end{equation}
or equivalently
\begin{equation}
\Omega_{m}(\mathbf{k})=-2\text{Im}\langle\hat{\mathbf{n}},m|\left[(\partial_{x}R)R^{\dagger},(\partial_{y}R)R^{\dagger}\right]|\hat{\mathbf{n}},m\rangle,\label{eq:Omega4}
\end{equation}
in the spin coherent basis (\ref{eq:coherent}), $|\hat{\mathbf{n}},m\rangle\equiv R(\hat{\mathbf{n}})|S,m\rangle$,
where we used $R(\partial_{x}R^{\dagger})=-(\partial_{x}R)R^{\dagger}$.
The infinitesimal spin rotation of $\hat{\mathbf{n}}\to\hat{\mathbf{n}}+\delta\hat{\mathbf{n}}$
is
\begin{equation}
R(\delta\hat{\mathbf{n}})=\text{exp}\left(i\hat{\mathbf{n}}\times\delta\hat{\mathbf{n}}\cdot\mathbf{S}\right),\label{eq:R}
\end{equation}
what yields
\begin{equation}
(\partial_{\gamma}R)R^{\dagger}=i\hat{\mathbf{n}}\times\partial_{\gamma}\hat{\mathbf{n}}\cdot\mathbf{S},\label{eq:Id}
\end{equation}
with $\gamma=x,y$. Thus, the Berry curvature of band $m$ is
\begin{align}
\Omega_{m} & =2\varepsilon_{ijk}(\partial_{x}\hat{\mathbf{n}}\times\hat{\mathbf{n}})_{i}(\partial_{y}\hat{\mathbf{n}}\times\hat{\mathbf{n}})_{j}\langle\hat{\mathbf{n}},m|S_{k}|\hat{\mathbf{n}},m\rangle.\nonumber \\
 & =2m\left(\partial_{x}\hat{\mathbf{n}}\times\partial_{y}\hat{\mathbf{n}}\right)\cdot\hat{\mathbf{n}},\label{eq:Omega3-1}
\end{align}
recovering Eq. (\ref{eq:Omega}). The quantum metric tensor of band
$m$ is defined as 
\begin{equation}
g_{m}^{\gamma\delta}=\langle S,m|(\partial_{\gamma}R^{\dagger})R(1-|S,m\rangle)R^{\dagger}(\partial_{\delta}R)|S,m\rangle.\label{eq:g3}
\end{equation}
Replacing Eq. (\ref{eq:Id}) in the expression above,
\begin{align*}
g_{m}^{\gamma\delta} & =\left(\partial_{\gamma}\hat{\mathbf{n}}\times\hat{\mathbf{n}}\right)_{i}\left(\partial_{\delta}\hat{\mathbf{n}}\times\hat{\mathbf{n}}\right)_{j}\\
 & \qquad\times\left[\langle S,m|S_{i}S_{j}|S,m\rangle-\delta_{ij}\delta_{i,3}\left(\langle S,m|S_{3}|S,m\rangle\right)^{2}\right],\\
 & =\frac{S(S+1)-m^{2}}{2}(\partial_{\gamma}\hat{\mathbf{n}})\cdot(\partial_{\delta}\hat{\mathbf{n}}),
\end{align*}
as stated in Eq. (\ref{eq:trace}). 

\section{Role of disorder}

Disorder introduces broadening to the fermionic propagators in the
current-current density correlations. In the weak disorder regime,
we can effectively introduce a fermionic broadening to the problem,
$\eta_{\text{f}}$, which we distinguish from the bosonic one $\eta_{\text{b}}$
originated from inelastic processes. The transport coefficient in
Eq. (\ref{eq:L}) can be directly expressed in terms of fermionic
Green's functions, \begin{widetext}
\begin{equation}
\text{Re}\left[L_{\gamma\delta}^{AB}(\omega)\right]=\frac{1}{4\pi\omega}\int_{-\infty}^{\infty}\text{d}\epsilon f(\epsilon)\sum_{\sigma=\pm}\text{Tr}\left\{ j_{\gamma}^{A}G(\epsilon+\sigma i\eta_{\text{f}})j_{\delta}^{B}\left[G(\epsilon+\omega+i\eta_{\sigma})-G(\epsilon-\omega+i\eta_{-\sigma})\right]\right\} +\text{h.c.},\label{eq:L3}
\end{equation}
\end{widetext}where $G_{\alpha\beta}(\epsilon+i\eta_{\text{f}})=(\epsilon-h_{\alpha\beta}+i\eta_{\text{f}})^{-1}$
is the retarded Green's function with fermionic broadening $\eta_{\text{f}}$
and $\eta_{\pm}=\eta_{\text{b}}\pm\eta_{\text{f}}$. Eq. (\ref{eq:L3})
recovers Eq. (\ref{eq:kubo_greenwood_wick-1})  in the clean limit
$\eta_{\text{f}}\to0$. In the DC limit, \begin{widetext}

\begin{equation}
L_{\gamma\delta}^{AB}(0)=\frac{1}{2\pi}\int_{-\infty}^{\infty}\text{d}\epsilon f(\epsilon)\text{Tr}\left\{ j_{\gamma}^{A}G(\epsilon+i\eta_{\text{f}})j_{\delta}^{B}\frac{\partial}{\partial\epsilon}G(\epsilon+i\eta_{+})-j_{\gamma}^{A}G(\epsilon-i\eta_{\text{f}})j_{\delta}^{B}\frac{\partial}{\partial\epsilon}G(\epsilon+i\eta_{-})\right\} +\text{h.c.}.\label{eq:L3-1}
\end{equation}
\end{widetext}One can further simplify the longitudinal part of the
transport coefficient tensor in the limit where $\eta_{\text{b}},\eta_{\text{f}}\to0$
and $\eta_{\text{b}}/\eta_{\text{f}}=\text{const}\gg1$, which reduces
to the standard Kubo-Streda formula when integrating by parts,
\begin{align}
L_{\gamma\gamma}^{AB}(0) & =\frac{1}{2\pi}\text{\ensuremath{\lim}}_{\eta_{\text{f}},\eta_{\text{b}}\to0}\int_{-\infty}^{\infty}\text{d}\epsilon\frac{\partial f(\epsilon)}{\partial\epsilon}\nonumber \\
 & \quad\times\text{Tr}\left\{ j_{\gamma}^{A}\text{Im}\left[G(\epsilon+i\eta_{\text{f}})\right]j_{\delta}^{B}\text{Im}\left[G^{R}(\epsilon+i\eta_{\text{b}})\right]\right\} .\label{eq:L4}
\end{align}
This formula should be used with caution, since $\eta_{\text{f}}$
and $\eta_{\text{b}}$ cannot be treated independently. A safer way
to include finite disorder would be to use Eq. (\ref{eq:L3-1}) instead,
where $\eta_{\text{f}}$ and $\eta_{\text{b}}$ can be treated as
independent parameters.

\end{document}